\documentclass[prl,twocolumn,a4paper,superscriptaddress,floatfix]{revtex4}

\usepackage{graphicx,amssymb}

\begin{document}
%My commands
\newcommand{\be}{\begin{equation}}
\newcommand{\ee}{\end{equation}}
\newcommand{\bq}{\begin{eqnarray}}
\newcommand{\eq}{\end{eqnarray}}
\newcommand{\bsq}{\begin{subequations}}
\newcommand{\esq}{\end{subequations}}
\newcommand{\bc}{\begin{center}}
\newcommand{\ec}{\end{center}}
\newcommand\lapp{\mathrel{\rlap{\lower4pt\hbox{\hskip1pt$\sim$}} \raise1pt\hbox{$<$}}}
\newcommand\gapp{\mathrel{\rlap{\lower4pt\hbox{\hskip1pt$\sim$}} \raise1pt\hbox{$>$}}}
\newcommand{\dpar}[2]{\frac{\partial #1}{\partial #2}}
\newcommand{\sdp}[2]{\frac{\partial ^2 #1}{\partial #2 ^2}}
\newcommand{\dtot}[2]{\frac{d #1}{d #2}}
\newcommand{\sdt}[2]{\frac{d ^2 #1}{d #2 ^2}}
\newcommand{\vv}{\bar{v}}

\title{Comment on ``Searching for Topological Defect Dark Matter via Nongravitational Signatures"}

\author{P.P. Avelino}
%\email[Electronic address: ]{pedro.avelino@astro.up.pt}
\affiliation{Instituto de Astrof\'{\i}sica e Ci\^encias do Espa{\c c}o, Universidade do Porto, CAUP, Rua das Estrelas, PT4150-762 Porto, Portugal}
\affiliation{Centro de Astrof\'{\i}sica da Universidade do Porto, Rua das Estrelas, PT4150-762 Porto, Portugal}
\affiliation{Departamento de F\'{\i}sica e Astronomia, Faculdade de Ci\^encias, Universidade do Porto, Rua do Campo Alegre 687, PT4169-007 Porto, Portugal}

\author{L. Sousa}
%\email[Electronic address: ]{Lara.Sousa@astro.up.pt}
\affiliation{Instituto de Astrof\'{\i}sica e Ci\^encias do Espa{\c c}o, Universidade do Porto, CAUP, Rua das Estrelas, PT4150-762 Porto, Portugal}
\affiliation{Centro de Astrof\'{\i}sica da Universidade do Porto, Rua das Estrelas, PT4150-762 Porto, Portugal}

\author{Francisco S. N. Lobo}
%\email{fslobo@fc.ul.pt}
\affiliation{Instituto de Astrof\'{\i}sica e Ci\^{e}ncias do Espa\c{c}o, Universidade de Lisboa, Faculdade de Ci\^{e}ncias, Campo Grande, PT1749-016 Lisboa, Portugal}
\affiliation{Faculdade de
Ci\^encias da Universidade de Lisboa, Edif\'{\i}cio C8, Campo Grande,
P-1749-016 Lisboa, Portugal}

\date{\today}

\begin{abstract}

In the letter by Stadnik and Flambaum [Phys. Rev. Lett. 113, 151301 (2014)] it is claimed that topological defects passing through pulsars could be responsible for the observed pulsar glitches. Here, we show that, independently of the detailed network dynamics and defect dimensionality, such proposal is faced with serious difficulties.

\end{abstract}

\maketitle

In \cite{Stadnik:2014cea}, the authors claimed that the pulsar glitch phenomenon \cite{Espinoza:2011pq,Haskell:2015jra,pulsar} might be caused by the passage of topological defects through pulsars. When a defect encounters a pulsar, its interaction with the neutrons may induce an increase of the neutron mass, and a release of a substantial fraction of the (local) defect kinetic energy. It has been argued in \cite{Stadnik:2014cea} that this energy-momentum  transfer might result in sudden changes in pulsar rotational periods, and thus explain the observed pulsar glitches. Here, we show that the maximum relative frequency change that could be induced by a defect in one glitch event is more than $15$ orders of magnitude smaller than the typical observed values. We also argue that, if the neutron mass is lowered by the defect, then the defect would become attached to the pulsar.

Let us start by treating the case in which neutron mass is increased by the defect passage and by considering a topological defect network with a local average density ${\rho}=\chi \rho_{\rm c}={3 \chi H^2}/{(8\pi G)}$.Here, $\rho_{\rm c}$ is the critical density, $H$ is the Hubble parameter, $G$ is the gravitational constant, $\chi > 0$, and all quantities are evaluated at the present time. This density may be used to define the characteristic lengthscale $L$ of the network, given by $\rho=\sigma_p L^ {p-3}$, where $\sigma_p$ is the defect mass per unit $p$-dimensional area, and $p$ is the defect dimensionality \cite{Sousa:2011ew,Sousa:2011iu} --- $p=0,1$ and $2$ for monopoles, strings and domain walls, respectively. When an encounter occurs, if the radius of the pulsar is $R\ll L$, only a small defect portion of area $\sim R^p$ may pass through the pulsar. The characteristic timescale $T$ between successive encounters of a defect with a given pulsar is thus the average time the pulsar would take to sweep a volume that contains on average a defect portion of $p$-dimensional area $\sim R^p$. $T$ may, then, be estimated by requiring the average defect mass contained in the volume swept by the pulsar (${\rho}R^2 vT$) to be approximately equal to the total mass inside a $p$-dimensional defect portion of area $R^p$: $\rho v T =\sigma_pR^{p-2}$. The maximum amount of energy available to be transferred from the defect to the pulsar may be estimated as the total kinetic energy inside a portion of the defect of area $r^p$, where $r=R c/v$ and $c$ is the speed of light in vacuum:
\bq
& & E_{\rm max}  \sim \sigma_p r^p v^2  \sim \nonumber \\
&\sim& 10^{-10+3p} \chi \frac{T}{\rm 10 \, yr} \left(\frac{R}{\rm 10 \, km}\right)^2 \left(\frac{v}{10^{-3} \, c}\right)^{3-p} {\rm kg \, c^2}\,. \nonumber
\eq
The typical timescale between successive glitch events is $T\sim 1\,-10 \, {\rm yr}$, and therefore we shall consider the most conservative scenario $T \sim \rm 10 \, yr$. Taking $M_{\rm pulsar} \sim M_\odot \sim 10^{30} \, {\rm kg}$ for the pulsar mass, and assuming  $R \sim \rm 10 \, km$, $v \sim 10^{-3} \, c$ and $\chi=1$ one obtains a fractional mass variation ${\delta M}/{M} \lapp {E_{\rm max}}/{(M_{\rm pulsar} c^2)} \sim 10^{3p-40}$, which is, even in the case of domain walls ($p=2$), significantly smaller than the range $\delta M /M \sim \delta \omega /\omega\sim10^{-11}-10^{-5}$ associated with pulsar glitches. Relaxing the above assumptions by assuming that the defects are semi-relativistic and taking a value of $\chi$ as large as $10^5$ --- which would (unrealistically) elevate the defect energy density within the galaxy to the same level as that of the dark matter ---, one obtains ${\delta M}/{M} \lapp {E_{\rm max}}/{(M_{\rm pulsar} c^2)} \lapp 10^{-26}$, for all $p$. This is more than $15$ orders of magnitude smaller than the typical energies associated to the observed glitches in the rotation of pulsars. Therefore, the defect does not have enough kinetic energy to overcome the repulsive potential barrier and enter the pulsar, unless the coupling between the defect and the neutrons is extremely weak.

On the other hand, if the neutron mass is lowered by the defect, the energy would initially flow from the pulsar into the defect. This implies that --- unless the coupling between the defect and neutrons is negligible --- the portion of the defect that enters the pulsar becomes highly relativistic. However, this local energy-momentum boost is expected to be dissipated very effectively, in particular due to the extremely fast energy-momentum flow towards the rest of the defect. As a consequence, the defect may not have enough kinetic energy to overcome the potential barrier and exit the pulsar, and could become attached to it. In this case, a rapid passage of the defect through the pulsar --- one of the necessary conditions for it to trigger a pulsar glitch --- would not occur.

%%%%%%%%%%%%%%%%%%%%%%%%%%%%%%%%%%%%%%%%%%%%%%%%%%%%%%%%%%
\acknowledgments{

P.P.A. and F.S.N.L. are supported by Funda{\c c}\~ao para a Ci\^encia e a Tecnologia (FCT, Portugal) FCT Research contracts of reference IF/00863/2012 and IF/00859/2012, respectively. L.S. is supported by FCT through the grant SFRH/BPD/76324/2011. Funding of this work was also provided by the FCT grant UID/FIS/04434/2013 and EXPL/FIS-AST/1608/2013.
}
%%%%%%%%%%%%%%%%%%%%%%%%%%%%%%%%%%%%%%%%%%%%%%%%%%%%%%%%%%

\bibliography{PRL_pulsar_glitches}

\begin{thebibliography}{6}
\expandafter\ifx\csname natexlab\endcsname\relax\def\natexlab#1{#1}\fi
\expandafter\ifx\csname bibnamefont\endcsname\relax
  \def\bibnamefont#1{#1}\fi
\expandafter\ifx\csname bibfnamefont\endcsname\relax
  \def\bibfnamefont#1{#1}\fi
\expandafter\ifx\csname citenamefont\endcsname\relax
  \def\citenamefont#1{#1}\fi
\expandafter\ifx\csname url\endcsname\relax
  \def\url#1{\texttt{#1}}\fi
\expandafter\ifx\csname urlprefix\endcsname\relax\def\urlprefix{URL }\fi
\providecommand{\bibinfo}[2]{#2}
\providecommand{\eprint}[2][]{\url{#2}}

\bibitem[{\citenamefont{Stadnik and Flambaum}(2014)}]{Stadnik:2014cea}
\bibinfo{author}{\bibfnamefont{Y.~V.} \bibnamefont{Stadnik}} \bibnamefont{and}
  \bibinfo{author}{\bibfnamefont{V.~V.} \bibnamefont{Flambaum}},
  \bibinfo{journal}{Phys.Rev.Lett.} \textbf{\bibinfo{volume}{113}},
  \bibinfo{pages}{151301} (\bibinfo{year}{2014}), \eprint{1405.5337}.

\bibitem[{\citenamefont{Espinoza et~al.}(2011)\citenamefont{Espinoza, Lyne,
  Stappers, and Kramer}}]{Espinoza:2011pq}
\bibinfo{author}{\bibfnamefont{C.~M.} \bibnamefont{Espinoza}},
  \bibinfo{author}{\bibfnamefont{A.~G.} \bibnamefont{Lyne}},
  \bibinfo{author}{\bibfnamefont{B.~W.} \bibnamefont{Stappers}},
  \bibnamefont{and} \bibinfo{author}{\bibfnamefont{M.}~\bibnamefont{Kramer}},
  \bibinfo{journal}{Mon.Not.Roy.Astron.Soc.} \textbf{\bibinfo{volume}{414}},
  \bibinfo{pages}{1679} (\bibinfo{year}{2011}), \eprint{1102.1743}.

\bibitem[{\citenamefont{Haskell and Melatos}(2015)}]{Haskell:2015jra}
\bibinfo{author}{\bibfnamefont{B.}~\bibnamefont{Haskell}} \bibnamefont{and}
  \bibinfo{author}{\bibfnamefont{A.}~\bibnamefont{Melatos}},
  \bibinfo{journal}{Int.J.Mod.Phys.} \textbf{\bibinfo{volume}{D24}},
  \bibinfo{pages}{1530008} (\bibinfo{year}{2015}), \eprint{1502.07062}.

\bibitem[{pul()}]{pulsar}
\emph{\bibinfo{title}{Atnf pulsar catalogue: Glitch parameters}},
  \urlprefix\url{http://www.atnf.csiro.au/people/pulsar/psrcat/glitchTbl.html}.

\bibitem[{\citenamefont{Sousa and Avelino}(2011{\natexlab{a}})}]{Sousa:2011ew}
\bibinfo{author}{\bibfnamefont{L.}~\bibnamefont{Sousa}} \bibnamefont{and}
  \bibinfo{author}{\bibfnamefont{P.~P.} \bibnamefont{Avelino}},
  \bibinfo{journal}{Phys. Rev.} \textbf{\bibinfo{volume}{D83}},
  \bibinfo{pages}{103507} (\bibinfo{year}{2011}{\natexlab{a}}),
  \eprint{1103.1381}.

\bibitem[{\citenamefont{Sousa and Avelino}(2011{\natexlab{b}})}]{Sousa:2011iu}
\bibinfo{author}{\bibfnamefont{L.}~\bibnamefont{Sousa}} \bibnamefont{and}
  \bibinfo{author}{\bibfnamefont{P.~P.} \bibnamefont{Avelino}},
  \bibinfo{journal}{Phys.Rev.} \textbf{\bibinfo{volume}{D84}},
  \bibinfo{pages}{063502} (\bibinfo{year}{2011}{\natexlab{b}}),
  \eprint{1107.4582}.

\end{thebibliography}

\end{document}